\documentclass[12pt,oneside,a4paper]{article}
\pdfoutput=1

\usepackage{hyperref}
\hypersetup{colorlinks,bookmarksopen,bookmarksnumbered,citecolor=blue,
linkcolor=black,pdfstartview=FitH,urlcolor=blue}

\usepackage{graphicx}
\usepackage{amssymb}
\usepackage{cite}
\usepackage{bm}
\usepackage{indentfirst}
\usepackage{amsmath}
\usepackage{hhline}
\usepackage{multirow}
\usepackage{color}

\allowdisplaybreaks
\numberwithin{equation}{section}

\def\Re{\mathop{\mathrm{Re}}}

\newcommand{\mbb}{\mathbb}

\definecolor{dred}{rgb}{0.7,0.15,0.09}
\definecolor{dblue}{rgb}{0,0.12,0.64}
\definecolor{dgreen}{rgb}{0.2,0.51,0.19}

\begin{document}

\begin{titlepage}

\begin{flushright}
KANAZAWA-21-09
\\
KUNS-2889
\end{flushright}

\begin{center}

\vspace{1cm}
{\large\textbf{
Direct detection of pseudo-Nambu-Goldstone dark matter\\ with light mediator
}
 }
\vspace{1cm}

\renewcommand{\thefootnote}{\fnsymbol{footnote}}
Yoshihiko Abe$^{1}$\footnote[1]{y.abe@gauge.scphys.kyoto-u.ac.jp}
and
Takashi Toma$^{2,3}$\footnote[2]{toma@staff.kanazawa-u.ac.jp}
\vspace{5mm}

\textit{
 $^1${Department of Physics, Kyoto University, Kyoto 606-8502, Japan}\\
 $^2${Institute of Liberal Arts and Science, Kanazawa University, Kanazawa, 920-1192 Japan}\\
 $^3${Institute for Theoretical Physics, Kanazawa University, Kanazawa 920-1192, Japan}
}

\vspace{8mm}

\abstract{
It has been found that a pseudo-Nambu-Goldstone boson dark matter suppresses the amplitude for elastic scattering with nuclei in non-relativistic limit, 
and thus can naturally evade the strong constraint of dark matter direct detection experiments. 
In this paper, we show that non-zero elastic scattering cross section can be induced 
if the mediator mass is as small as momentum transfer.
The predicted recoil energy spectrum can differ from that for usual thermal dark matter. 
Together with the relevant constraints such as thermal relic abundance, indirect detection and Higgs decays, 
we investigate the detectability through the current and future dark matter direct detection experiments. 
}

\end{center}
\end{titlepage}

\renewcommand{\thefootnote}{\arabic{footnote}}
\newcommand{\bhline}[1]{\noalign{\hrule height #1}}
\newcommand{\bvline}[1]{\vrule width #1}

\setcounter{footnote}{0}

\setcounter{page}{1}

\section{Introduction}

Dark matter direct detection experiments impose strong bounds on thermal dark matter. 
The current strongest bound is given by the XENON1T/PandaX-4T~\cite{XENON:2018voc, PandaX-4T:2021bab}, 
and will be further updated by the future XENONnT experiment~\cite{XENON:2020kmp}.
The pseudo-Nambu-Goldstone boson (pNGB) dark matter has been proposed as a candidate naturally evading this strong constraint~\cite{Gross:2017dan}. 
This is because the amplitude of the elastic scattering with nuclei is suppressed by the small momentum of dark matter through the derivative coupling. 
A non-zero elastic cross section is induced at loop levels, however it is marginally small compared to future sensitivities~\cite{Azevedo:2018exj, Ishiwata:2018sdi, Glaus:2020ihj}. 
In addition, a sophisticated global fit of the model with relevant constraints has also been performed~\cite{Arina:2019tib}.

For the pNGB dark matter model, an explicit soft breaking term is required to give a mass for the NGB after the spontaneous breaking of the corresponding global symmetry. 
It may be natural that the scale of the soft breaking term is much smaller than the scale of the spontaneous symmetry breaking in the sense that the global symmetry is approximate. 
However this is not necessarily true if a ultra-violet (UV) completion of the pNGB model is considered~\cite{Abe:2020iph, Okada:2020zxo, Abe:2021byq, Okada:2021qmi}. 
In this case, the effective pNGB model derived at low energy can have a soft breaking term larger than the scale of the spontaneous symmetry breaking. 
As a result, the mass of the particle mediating between dark matter and Standard Model (SM) particles can be much lighter than the dark matter mass.

In this paper, we consider the pNGB dark matter with a light mediator. 
In this scenario, we will show that non-zero elastic scattering cross section between dark matter and nuclei emerges even though the interactions arise from the derivative coupling. 
Furthermore, this scenario has a potential to discriminate the pNGB dark matter and the other dark matter candidates such as Weakly Interacting Massive Particles (WIMPs) 
by comparing the recoil energy spectrum of the event rate at direct detection. 
We contemplate some relevant constraints such as thermal production of dark matter, indirect detection bounds and Higgs invisible decay. 
Then, we show that some parameter region has already been excluded by the XENON1T bound, and some other region will be explored by the future XENONnT experiment.

\section{The model}
We consider the model extended with a complex singlet scalar $S$. 
The model has a global $U(1)_S$ symmetry under the transformation $S\to e^{i\alpha}S$, which is broken to the $\mathbb{Z}_2$ parity by introducing the soft breaking term $S^2$. 
The soft breaking term can be derived from a UV completion of the model~\cite{Abe:2020iph, Okada:2020zxo, Abe:2021byq, Okada:2021qmi}. 
The Lagrangian is given by
\begin{align}
 \mathcal{L}=\left|\partial_{\mu}S\right|^2-\mathcal{V}(H,S),
\end{align}
where the scalar potential $\mathcal{V}$ is written down as
\begin{equation}
 \mathcal{V}=-\frac{\mu_H^2}{2}|H|^2-\frac{\mu_S^2}{2}|S|^2+\frac{\lambda_H}{2}|H|^4+\lambda_{HS}|H|^2|S|^2+\frac{\lambda_S}{2}|S|^4-\frac{m_\chi^2}{4}\left(S^2+{S^*}^2\right).
\end{equation}
The $\lambda_{HS}$ term is the Higgs portal coupling giving the interaction between the SM and $S$.
The last term is the soft breaking term giving the mass to the NGB after the spontaneous symmetry breaking. 
Then, the scalar fields $H$ and $S$ can be parametrized as
\begin{equation}
 H=\frac{1}{\sqrt{2}}
\begin{pmatrix}
 0\\
v+h
\end{pmatrix},\quad
S=\frac{1}{\sqrt{2}}\left(v_s+s\right)e^{i\chi/v_s},
\label{eq:hs}
\end{equation}
where $v,v_s \in \mbb{R}$ are the vacuum expectation values
for $H$ and $S$, respectively.
Because of the interests in the low energy dynamics of pNGB and the simplicity of calculation,
we use the non-linear representation for $S$ in this paper.
These expectation values satisfy the following stationary conditions:
\begin{align}
 \mu_H^2 = \lambda_H v^2 + \lambda_{HS} v_s^2,
 \quad
 \mu_S^2 + m_\chi^2 = \lambda_S v_s^2 + \lambda_{HS} v^2.
 \label{eq:stationary}
\end{align}
The CP-even components $h$ and $s$ mix with each other via the Higgs portal coupling, and the mass matrix is given by
\begin{align}
\begin{pmatrix}
 \lambda_Hv^2 & \lambda_{HS}vv_s\\
\lambda_{HS}vv_s & \lambda_Sv_s^2
\end{pmatrix}.
\end{align}
This mass matrix is diagonalized by the unitary matrix, and thus the gauge eigenstates $h$ and $s$ can be rewritten by the mass eigenstates $h_1$ and $h_2$ as
\begin{align}
\begin{pmatrix}
 h\\
 s
\end{pmatrix}=
\begin{pmatrix}
 \cos\theta & \sin\theta\\
-\sin\theta & \cos\theta
\end{pmatrix}
\begin{pmatrix}
 h_1\\
h_2
\end{pmatrix},
\end{align}
where $h_1$ is identified as the SM-like Higgs boson with the mass $m_{h_1}=125~\mathrm{GeV}$ and $h_2$ is the second Higgs boson whose mass is assumed to be much lighter than $m_{h_1}$.
The CP-odd component $\chi$ is the pNGB with the mass $m_\chi$ and can be
a stable dark matter candidate due to
a $\mbb{Z}_2$ parity associated with the CP symmetry in the dark sector. 
The dark matter mass is given by the soft breaking parameter $m_\chi$. 

The quartic couplings $\lambda_H$, $\lambda_{HS}$ and $\lambda_S$ in the scalar potential can be rewritten in terms of the physical quantities as
\begin{align}
 \lambda_H&=\frac{\cos^2\theta m_{h_1}^2 + \sin^2\theta m_{h_2}^2}{v^2},\\
 \lambda_{HS}&=\frac{\sin\theta\cos\theta\left(m_{h_2}^2-m_{h_1}^2\right)}{vv_s},\\
 \lambda_S&=\frac{\sin^2\theta m_{h_1}^2 + \cos^2\theta m_{h_2}^2}{v_s^2},
\end{align}
and the cubic couplings which are relevant to the subsequent sections are given by
\begin{align}
\kappa_{111}&=3m_{h_1}^2\left(-\frac{\sin^3\theta}{v_s}+\frac{\cos^3\theta}{v}\right),
\label{eq:kappa1}\\
\kappa_{112}&=\left(2m_{h_1}^2+m_{h_2}^2\right)\sin\theta\cos\theta\left(\frac{\sin\theta}{v_s}+\frac{\cos\theta}{v}\right),
\label{eq:kappa2}\\
\kappa_{122}&=\left(m_{h_1}^2+2m_{h_2}^2\right)\sin\theta\cos\theta\left(-\frac{\cos\theta}{v_s}+\frac{\sin\theta}{v}\right),
\label{eq:kappa3}\\
\kappa_{222}&=3m_{h_2}^2\left(\frac{\cos^3\theta}{v_s}+\frac{\sin^3\theta}{v}\right),
\label{eq:kappa4}
\end{align}
with the convention
\begin{align}
\mathcal{V}\supset \frac{\kappa_{111}}{3!}h_1^3+\frac{\kappa_{112}}{2!}h_1^2h_2+\frac{\kappa_{122}}{2!}h_1h_2^2+\frac{\kappa_{222}}{3!}h_2^3.
\end{align}
Note that the interactions between the dark matter and the SM particles
come from the kinetic term $|\partial_{\mu}S|^2$ and the soft breaking term $ -\frac{m_\chi^2}{4} ( S^2 + {S^*}^2)$
since the field $S$ is written in non-linear representation as Eq.~(\ref{eq:hs}).

It could be natural from the 't Hooft sense that the soft breaking parameter $m_\chi$ is much smaller than the scale of the spontaneous symmetry breaking $v_s$ 
because the $\mathbb{Z}_2$ parity is enhanced to the global $U(1)_S$ symmetry 
in $m_\chi\to 0$ limit. 
However note that $m_\chi$ and $v_s$ are irrelevant and can be regarded as independent parameters once a UV completion of the model is considered~\cite{Abe:2020iph, Abe:2021byq, Okada:2020zxo, Okada:2021qmi}. 
Therefore, in the following we consider the case of $v_s,m_{h_2}\ll m_\chi$. 
From Eq.~\eqref{eq:stationary} and $m_{h_2}^2 \approx \lambda_S v_s^2$,
it is found that the quadratic mass parameter $\mu_S^2$ should be negative and $m_\chi^2$ is the trigger of the symmetry breaking in this case.
Although the breaking pattern is different,
the interactions among the scalars do not change from the ordinal pNGB dark matter model due to $v_s \in \mbb{R}$~\cite{Gross:2017dan}.
However, in the light mediator case ($m_{h_2} < m_\chi$),
we will show that non-zero elastic cross section with nuclei is induced and can be tested by the current and future direct detection experiments.

In particular, we focus on the mediator mass range of $20~\mathrm{MeV}\lesssim m_{h_2}\lesssim 200~\mathrm{MeV}$. 
This is because the lighter mass region has already been ruled out by the constraint of big bang nucleosyntheis (BBN)~\cite{Fradette:2017sdd} 
while the heavier mass region is insensitive to dark matter direct detection as will be seen below. 
In the mass range we focus on, the mixing angle $\sin\theta$ is roughly constrained in the range of $2\times10^{-5}\lesssim\sin\theta\lesssim10^{-4}$~\cite{Winkler:2018qyg}. 
The upper and lower bounds come from the constraints of the meson decays and BBN, respectively.

\section{The constraints}
Before proceeding to the main subject, we review the constraints relevant to our dark matter scenario. 

\subsection{Relic abundance}

\begin{figure}[t]
\begin{center}
\includegraphics[scale=0.85]{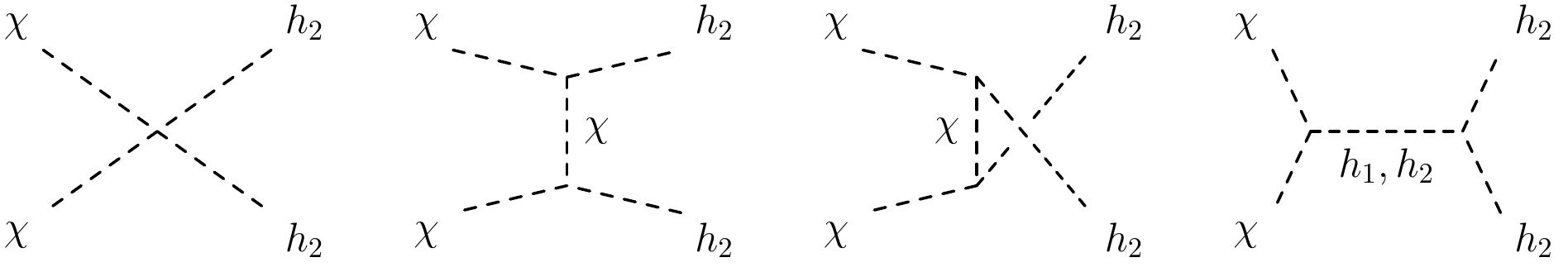}
\caption{Diagrams for the main dark matter annihilation $\chi\chi\to h_2h_2$.}
\label{fig:ann}
\end{center}
\end{figure}

In the current scenario, the dark matter annihilation channel $\chi\chi\to h_2h_2$ is dominant in the most of the parameter space. 
The cross section is calculated from the four diagrams shown in Fig.~\ref{fig:ann} as 
\begin{align}
\sigma_{h_2h_2}v_\mathrm{rel}\approx
\frac{\lambda_S^2}{16 \pi s}\left|1 + \frac{m_{h_1}^2}{m_{h_2}^2} \frac{ \sin^2 \theta s}{s - m_{h_1}^2 + im_{h_1} \Gamma_{h_1}} \right|^2, 
\label{eq:h2h2}
\end{align}
where $s$ denotes the Mandelstam variable. 
In the above, $\sin\theta\ll1$, $m_{h_2}\ll m_{h_1},m_\chi$ and $m_{h_2}^2\approx\lambda_Sv_s^2$ are used to simplify the equation.

For additional channels $\chi\chi\to WW,ZZ,f\overline{f}$, the cross sections are calculated as
\begin{align}
\sigma_{WW}v_\mathrm{rel}&\approx
\frac{\lambda_S g_2^2m_W^2}{8\pi}\sqrt{1-\frac{4m_W^2}{s}}
\frac{\sin^2\theta}{sm_{h_2}^2}
\left[3-\frac{s}{m_W^2}+\frac{s^2}{4m_W^4}\right]
\frac{m_{h_1}^4}{(s-m_{h_1}^2)^2+m_{h_1}^2\Gamma_{h_1}^2},
\label{eq:ww}\\
\sigma_{ZZ}v_\mathrm{rel}&\approx
\frac{\lambda_S g_2^2m_Z^2}{16\pi \cos^2\theta_W}\sqrt{1-\frac{4m_Z^2}{s}}
\frac{\sin^2\theta}{sm_{h_2}^2}
\left[3-\frac{s}{m_Z^2}+\frac{s^2}{4m_Z^4}\right]
\frac{m_{h_1}^4}{(s-m_{h_1}^2)^2+m_{h_1}^2\Gamma_{h_1}^2},
\label{eq:zz}\\
\sigma_{f\overline{f}}v_\mathrm{rel}&\approx
\frac{\lambda_Sg_2^2m_f^2}{32\pi}
\left(1-\frac{4m_f^2}{s}\right)^{3/2}
\frac{\sin^2\theta}{m_W^2m_{h_2}^2}
\frac{m_{h_1}^4}{(s-m_{h_1}^2)^2+m_{h_1}^2\Gamma_{h_1}^2},
\label{eq:ff}
\end{align}
where $g_2$ is the gauge coupling of the $SU(2)_L$ in the SM and $f\overline{f}$ represents a pair of the SM fermions. 
Although the other channels $\chi\chi\to h_1h_1,h_1h_2$ are also possible, the expressions are complicated and not shown here. 
The full expressions of the annihilation cross sections are given in Appendix~\ref{app:cross-section}. 
Note that all the cross sections except for the main channel $\chi\chi\to h_2h_2$ are suppressed by the small mixing angle $\sin^2\theta$. 
Thus it can be seen that the channel $\chi\chi\to h_2h_2$ tends to be dominant. 
The annihilation cross section for $\chi\chi\to h_2h_2$ in Eq.~(\ref{eq:h2h2}) is proportional to $\lambda_S^2$ 
while those for the other channels in 
Eqs.~(\ref{eq:ww})--(\ref{eq:ff}) are proportional to $\lambda_S$.

Then, the thermal averaged cross section can be given by~\cite{Gondolo:1990dk}
\begin{align}
 \langle\sigma{v}_\mathrm{rel}\rangle=\frac{1}{16m_\chi^4TK_2^2(m_\chi/T)}
\int_{4m_\chi^2}^{\infty}
\left(\sigma v_\mathrm{rel}\right)\sqrt{s-4m_\chi^2}sK_1\left(\frac{\sqrt{s}}{T}\right)
ds,
\end{align}
where $T$ is the temperature of the universe and $K_n(z)~(n=1,2)$ denotes the second kind modified Bessel function.
The Boltzmann equation is numerically solved by micrOMEGAs~\cite{Belanger:2018ccd} with the above analytic formulas for the annihilation cross sections, 
and the relic abundance should accommodate the PLANCK observation $\Omega_\chi h^2\approx0.12$~\cite{Planck:2018vyg}.\footnote{In the analytic calculations of the cross sections, a large cancellation between the corresponding diagrams occurs due to smallness of $v_s$. This causes huge numerical errors if the cross sections are automatically evaluated by using public codes such as CalcHEP~\cite{Belyaev:2012qa}. 
Thus the above analytic results are used to evaluate the thermal relic abundance of dark matter in micrOMEGAs~\cite{Belanger:2018ccd}. 
} 
Note that the early kinetic decoupling effect may change the parameter space which can reproduce the observed relic abundance when the dark matter mass is close to the SM-like Higgs resonance~\cite{Abe:2021jcz}.

\subsection{Indirect detection}
The main annihilation channel $\chi\chi\to h_2h_2$ generates cosmic rays in the galaxy via subsequent $h_2$ decays. 
In the mass range we focus on ($30~\mathrm{MeV}\lesssim m_{h_2}\lesssim 200~\mathrm{MeV}$), 
the main decay channel is $h_2\to e^+e^-$, which is constrained by the $e^+e^-$ observation of AMS-02~\cite{AMS:2014bun}. 
In addition to the $e^+e^-$ flux, gamma rays are also generated via final state radiation from the produced $e^+e^-$ 
and bounded by the observations from dwarf spheroidal galaxies at Fermi-LAT~\cite{Fermi-LAT:2016uux}. 
However this constraint is weaker than the $e^+e^-$ production~\cite{Elor:2015bho}.\footnote{The other annihilation channels into the SM particles also produce gamma rays. However the constraint is weaker than that of $e^+e^-$ due to the small cross sections.}
The Cosmic Microwave Background (CMB) may also be distorted by the dark matter annihilation 
because the produced charged particles and gamma rays ionize the universe after the recombination era. 
Thus the CMB measurement by PLANCK also sets a bound on the model.

\begin{figure}[t]
\begin{center}
\includegraphics[scale=0.65]{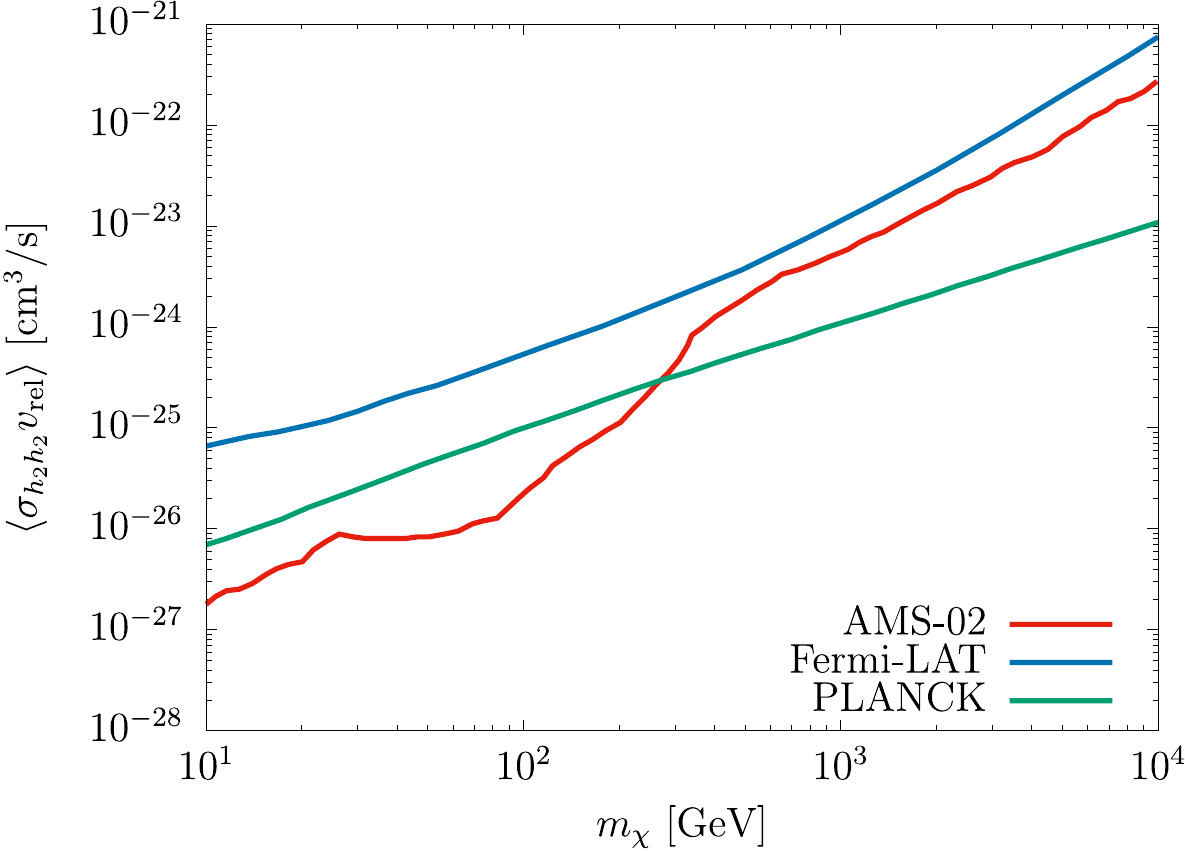}
\caption{Bounds on the annihilation cross section for $\chi\chi\to h_2h_2$ from indirect detection extracted from the literature~\cite{Elor:2015bho}.}
\label{fig:sigmav_limit} 
\end{center}
\end{figure}

We adopt the model independent bounds for multi-step cascade decays derived in the literature~\cite{Elor:2015bho} (top left panel in Fig.~11 therein). 
The extracted bounds are shown in Fig.~\ref{fig:sigmav_limit}.
As can be seen, the AMS-02 bound is severe for $m_{\chi}\lesssim100~\mathrm{GeV}$. 
For $m_\chi\gtrsim300~\mathrm{GeV}$, the CMB bound can be stronger than the AMS-02 bound. 
However the bound requires $\langle\sigma{v}_\mathrm{rel}\rangle\lesssim\mathcal{O}(10^{-25})~\mathrm{cm^3/s}$, which is consistent with thermal relic abundance. 

Another comment is that the light mediator $h_2$ does not induce a long range interaction. 
Thus we do not need to care about the non-perturbative Sommerfeld effects for the dark matter annihilation~\cite{Hisano:2002fk, Hisano:2003ec, Hisano:2004ds, Hisano:2005ec, Hisano:2006nn}.
This is due to the nature of the pNGB dark matter that all the interactions are written by the derivatives couplings.

\subsection{Higgs decay}
Since the mediator $h_2$ is light enough, the SM-like Higgs boson $h_1$ decays into $h_2h_2$ whose decay width is calculated as
\begin{align}
\Gamma_{h_1\to h_2h_2}=\frac{\kappa_{122}^2}{32\pi m_{h_1}}\sqrt{1-4\frac{m_{h_2}^2}{m_{h_1}^2}}
\approx
\frac{\lambda_Sm_{h_1}^3\sin^2\theta}{32\pi m_{h_2}^2},
\label{eq:hdecay}
\end{align} 
where $\sin\theta\ll1$ and $m_{h_2}\ll m_{h_1}$ are assumed and $m_{h_2}^2\approx\lambda_Sv_s^2$ is used.
This decay channel can be regarded as an invisible decay because the lifetime of $h_2$ is long enough to escape the detector of colliders. 
In addition to this decay channel, another channel $h_1\to\chi\chi$ is also possible if the dark matter mass is $m_\chi< m_{h_1}/2\approx 62.5~\mathrm{GeV}$.
The decay width is evaluated as
\begin{align}
\Gamma_{h_1\to\chi\chi}=
\frac{m_{h_1}^3\sin^2\theta}{32\pi v_s^2}\sqrt{1-4\frac{m_\chi^2}{m_{h_1}^2}}
\approx
\frac{\lambda_Sm_{h_1}^3\sin^2\theta}{32\pi m_{h_2}^2}\sqrt{1-4\frac{m_\chi^2}{m_{h_1}^2}}.
\label{eq:hdecay2}
\end{align}
The observation of the Higgs signal strength at the LHC is translated into the constraint of the Higgs invisible decay whose branching fraction should satisfy $\mathrm{Br}_\mathrm{inv}\leq 0.11$~\cite{CMS:2018yfx, ATLAS:2019cid}.

\section{Direct detection of pNGB dark matter}

It is known that the amplitude for the elastic scattering between the pNGB dark matter and a nucleus $\chi A\to \chi A$ vanishes in non-relativistic limit of dark matter. 
This argument is based on the premise that the mass of the mediator $h_2$ is much heavier than the momentum transfer $q=\sqrt{2m_A E_R}$~\cite{Gross:2017dan}. 
However this is not the case in our scenario where the mass of the mediator $h_2$ can be the same order or smaller than the momentum transfer.
Assuming $q^2,m_{h_2}^2\ll m_{h_1}^2$ and $\sin\theta\ll1$, the differential cross section for the elastic scattering with a nucleus $A$ is calculated as
\begin{align}
\frac{d\sigma_A}{dE_R}
\approx
\frac{m_A\kappa_{A}^2}{8\pi m_\chi^2 v_\chi^2}
\frac{\sin^2\theta}{v^2v_s^2}
\frac{t^2}{\left(t-m_{h_2}^2\right)^2}F^2(E_R),
\label{eq:dsde}
\end{align}
where $t = -q^2$ is the Mandelstam variable, $v_\chi$ is the velocity of dark matter and $F(E_R)$ is the nuclear Helm form factor which is parametrized by~\cite{Lewin:1995rx}
\begin{align}
F(E_R)=\frac{3j_1(qR)}{qR}e^{-q^2\tilde{s}^2/2},
\end{align}
with the spherical Bessel function $j_1(qR)$, $R\approx\sqrt{r^2-5\tilde{s}^2}~\mathrm{fm}$, $r\approx1.2A^{1/3}~\mathrm{fm}$, 
($A$ is the mass number of the target nucleus) and $\tilde{s}\approx1~\mathrm{fm}$. 
The Helm form factor is normalized as $F(0)=1$. 
The coupling $\kappa_A$ in Eq.~(\ref{eq:dsde}) is given by
\begin{align}
 \kappa_A=\kappa_pZ + \kappa_n(A-Z)
=\sum_{q}\Big[
f_q^p m_pZ + f_q^n m_n(A-Z)
\Big],
\end{align}
where $A$ and $Z$ are the mass number and the atomic number of the nucleus, respectively. 
The coefficients $f_q^p$ and $f_q^n$ are the scalar quark form factors of a nucleon, which are chosen as the values in the literature~\cite{Belanger:2018ccd} 
\begin{align}
&f_{u}^p=0.0153,\quad f_{d}^p=0.0191,\quad f_{s}^p=0.0447,\nonumber\\
&f_{u}^n=0.0110,\quad f_{d}^n=0.0273,\quad f_{s}^n=0.0447,
\end{align}
for the light quarks. The form factors of the heavy quarks can be written in terms of those of the light quarks. 

Direct detection experiments provide the limits on the elastic scattering cross section with a proton at zero momentum transfer, and 
the current strongest limit is given by XENON1T/PandaX-4T experiments~\cite{XENON:2018voc, PandaX-4T:2021bab}. 
Note that this limit cannot directly be applied to the current scenario because one cannot simply take zero momentum transfer limit in the differential cross section in Eq.~(\ref{eq:dsde}) 
due to a strong dependence on the momentum transfer and light mediator mass.
However, the experimental limit can be translated into the limit on the total event rate. 
Then, it is compared with the predicted event rate at a given parameters in the model.\footnote{A similar approach for direct detection with long range interactions due to a light mediator has been discussed~\cite{Hambye:2018dpi}.} 
The differential event rate is given by
\begin{align}
\frac{dR}{dE_R}=\frac{\rho_\odot}{m_\chi}N_T\int_{v_\chi>v_\mathrm{min}}\frac{d\sigma_A}{dE_R}v_\chi f_\odot(\bm{v}_\chi)d^3v_\chi,
\end{align}
where $\rho_\odot\approx0.3~\mathrm{GeV/cm^3}$ is the local dark matter density, $N_T$ is the number of target nucleus, 
$f_\odot(\bm{v}_\chi)$ is the Maxwell-Boltzmann velocity distribution function at the solar system~\cite{McCabe:2010zh}, and 
$v_\mathrm{min}$ is the minimum velocity of dark matter at the given recoil energy $E_R$ 
\begin{align}
v_\mathrm{min}=\sqrt{\frac{m_AE_R}{2}}\frac{1}{\mu_{A\chi}},
\end{align}
where $\mu_{A\chi}=m_Am_\chi/(m_A+m_\chi)$ is the reduced mass between dark matter and nucleus.

\begin{figure}[t]
\centering
\includegraphics[scale=0.65]{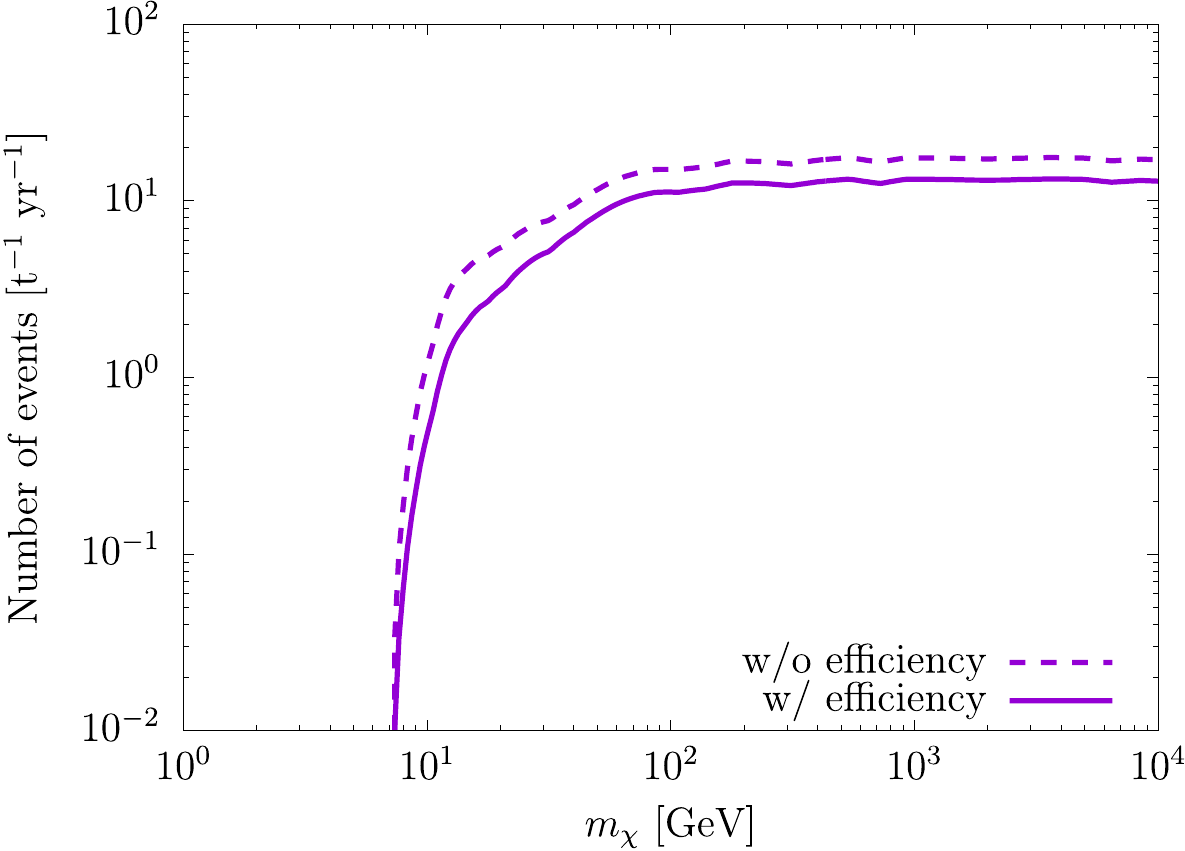}
\includegraphics[scale=0.65]{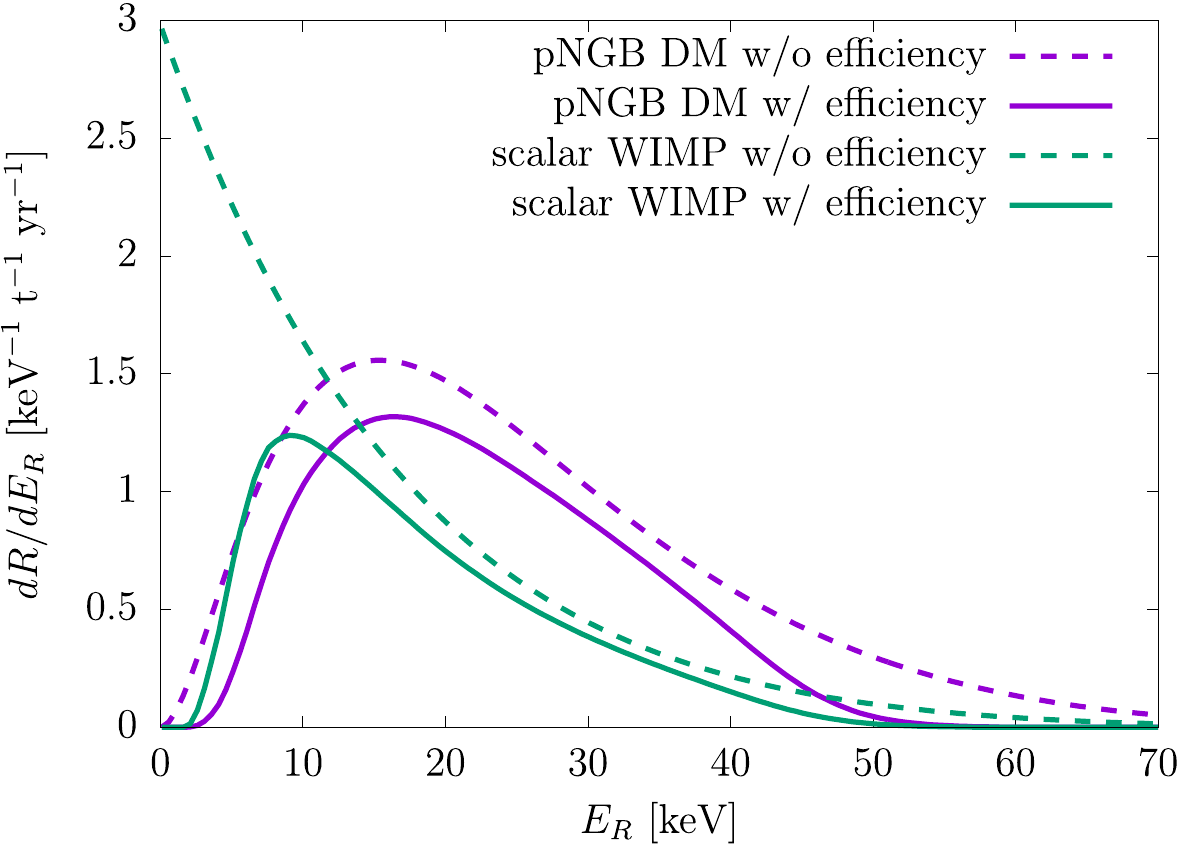}
\caption{(Left): Limit on the total event rate translated from the XENON1T experiment in the range of the recoil energy $4.9~\mathrm{keV}<E_R<40.9~\mathrm{keV}$ 
and $1.0$ t$\times$yr exposure~\cite{XENON:2018voc}. The solid (dashed) line represent the limit with (without) the efficiency of the XENON1T detector.
(Right): Comparison of energy spectra of number of events for pNGB dark matter and singlet scalar dark matter with $1.0$ t$\times$yr exposure 
where dark matter mass is fixed to be $m_{\chi}=150~\mathrm{GeV}$. 
The other parameters are taken as $\lambda_S=0.09$, $\sin\theta=3\times10^{-5}$ and $m_{h_2}=60~\mathrm{MeV}$ for the pNGB dark matter, 
and $\lambda_{hS}=0.006$ for the singlet scalar dark matter~\cite{GAMBIT:2017gge}. 
The solid (dashed) lines represent the spectra with (without) detector efficiency (XENON1T)~\cite{XENON:2018voc}.
}
\label{fig:dd0} 
\end{figure}

For the experimental limit, assuming no isospin violation~\cite{Feng:2011vu}, 
one can parametrize the differential cross section between dark matter and a nucleus as
\begin{align}
 \frac{d\sigma_A^\mathrm{exp}}{dE_R}=\frac{m_AA^2}{2\mu_{p\chi}^2v_\chi^2}\sigma_p^\mathrm{exp}F^2(E_R),
\end{align}
where $\sigma_p^\mathrm{exp}$ is the experimental upper limit on the total elastic cross section with a proton. 
Using this parametrization, the differential event rate with the experimental limit $\sigma_p^\mathrm{exp}$ can be rewritten as 
\begin{align}
\frac{dR^\mathrm{exp}}{dE_R}=\frac{\rho_\odot}{m_\chi}N_T\frac{m_AA^2}{2\mu_{p\chi}^2}\sigma_p^\mathrm{exp}F^2(E_R)
\int_{v_\chi>v_\mathrm{min}}\frac{f_\odot(\bm{v}_\chi)}{v_\chi}d^3v_\chi.
\end{align}
The left panel of Fig.~\ref{fig:dd0} shows the upper bound on the total event rate $R^\mathrm{exp}$ obtained from the XENON1T limit on 
the total elastic scattering cross section $\sigma_p^\mathrm{exp}$. 
The interest of region for the recoil energy is $4.9~\mathrm{keV}<E_R<40.9~\mathrm{keV}$ and the exposure $1.0~\mathrm{t}\times\mathrm{yr}$ is taken into account~\cite{XENON:2018voc}.

\begin{figure}[t]
\centering
  \includegraphics[scale=0.65]{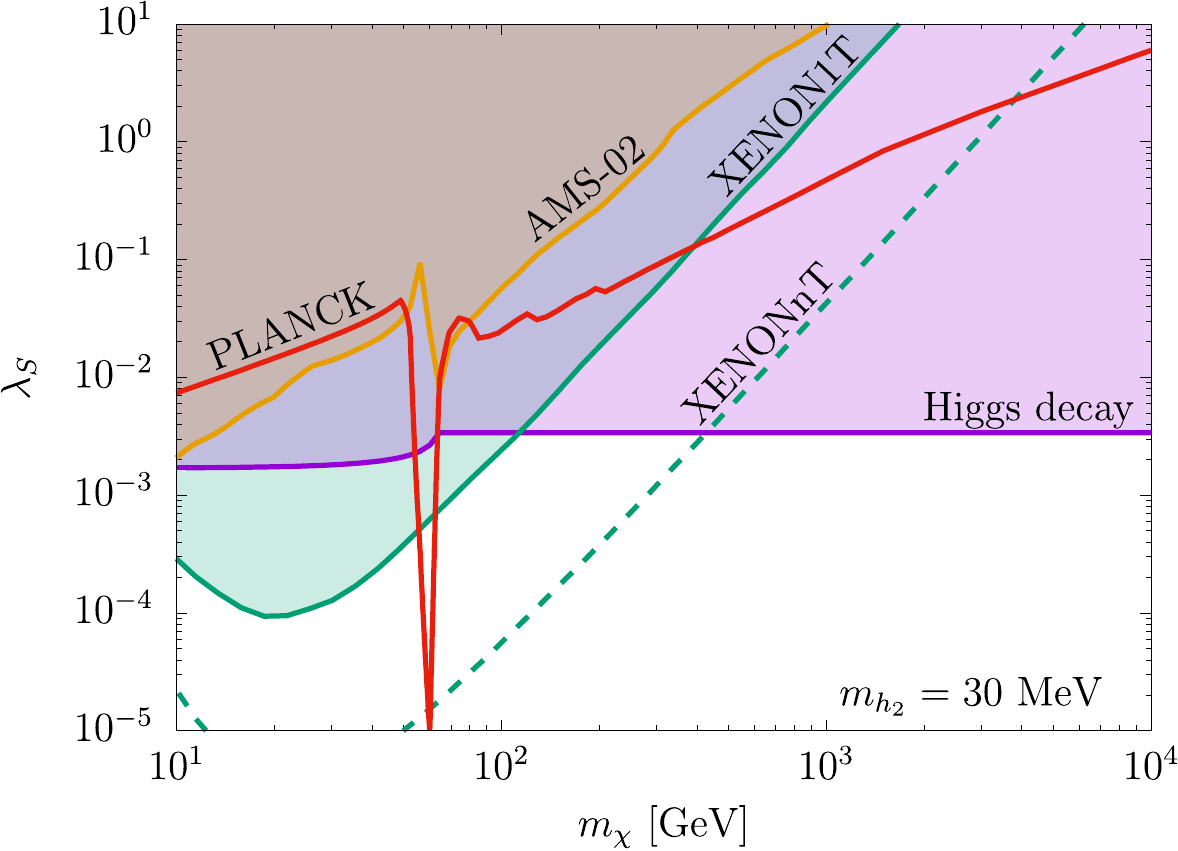}
  \includegraphics[scale=0.65]{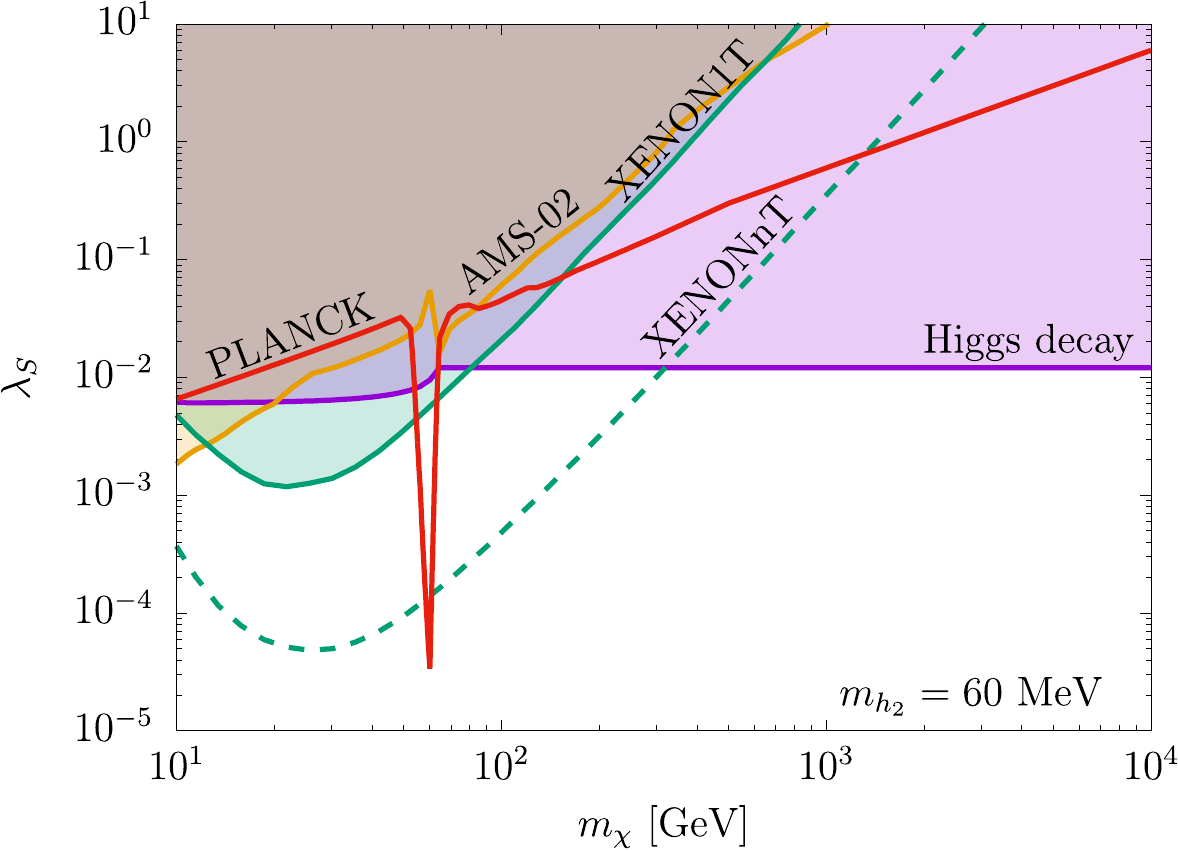}\\
  \includegraphics[scale=0.65]{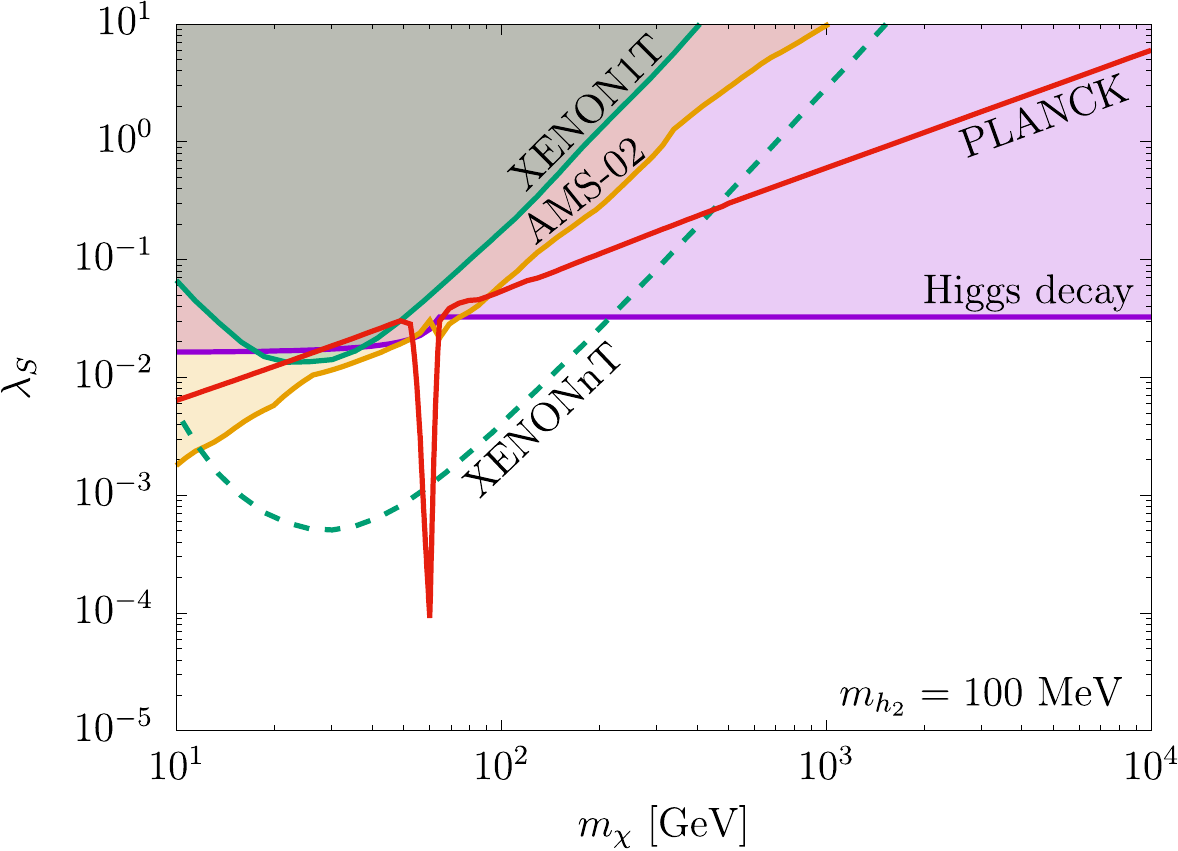}
  \includegraphics[scale=0.65]{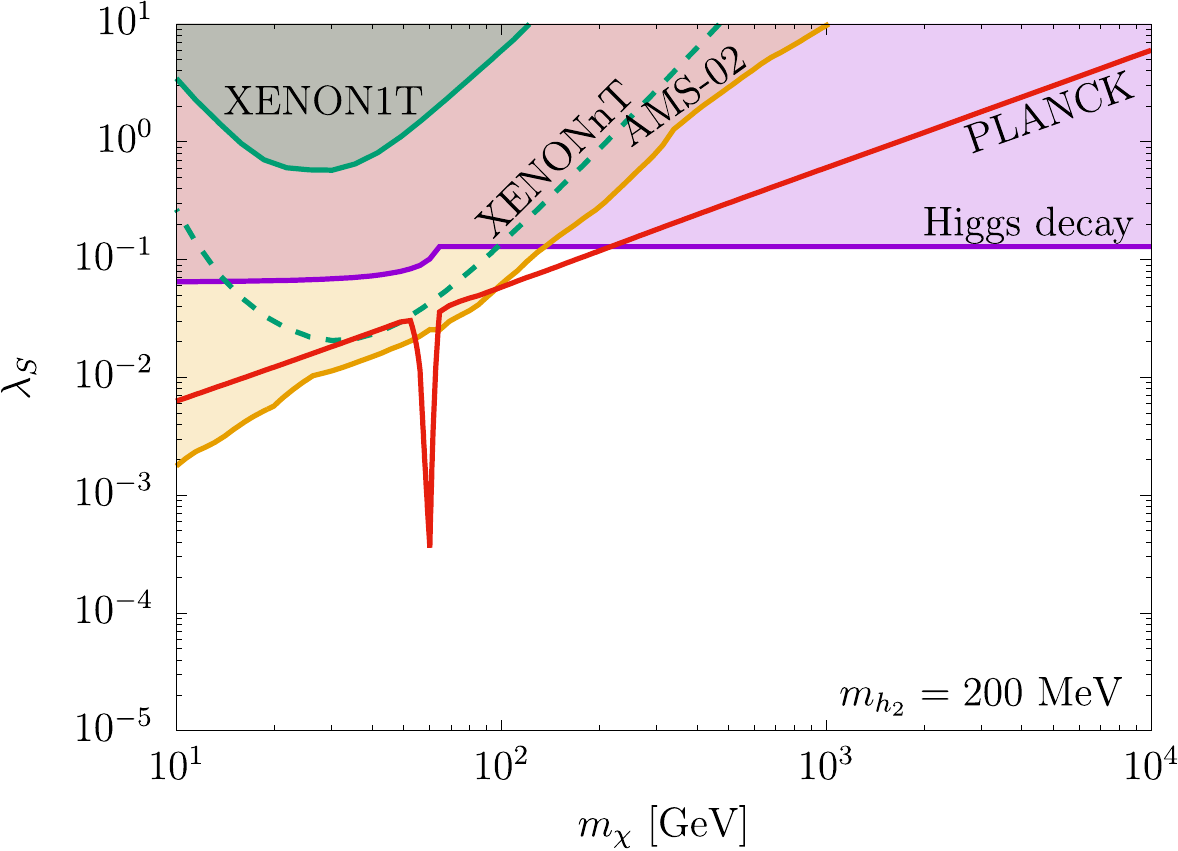}
\caption{Parameter region in the ($m_\chi$, $\lambda_S$) plane for $\sin\theta=10^{-4}$ and $m_{h_2}=30,60,100$ and $200~\mathrm{MeV}$. 
The green, orange and violet region are excluded by XENON1T~\cite{XENON:2018voc}, AMS-02~\cite{AMS:2014bun} and Higgs decays~\cite{CMS:2018yfx, ATLAS:2019cid}, respectively. 
The red line corresponds to the parameter space which can reproduce the observed relic abundance by PLANCK~\cite{Planck:2018vyg}. }
\label{fig:dd}
\end{figure}

\begin{figure}[t]
\centering
  \includegraphics[scale=0.65]{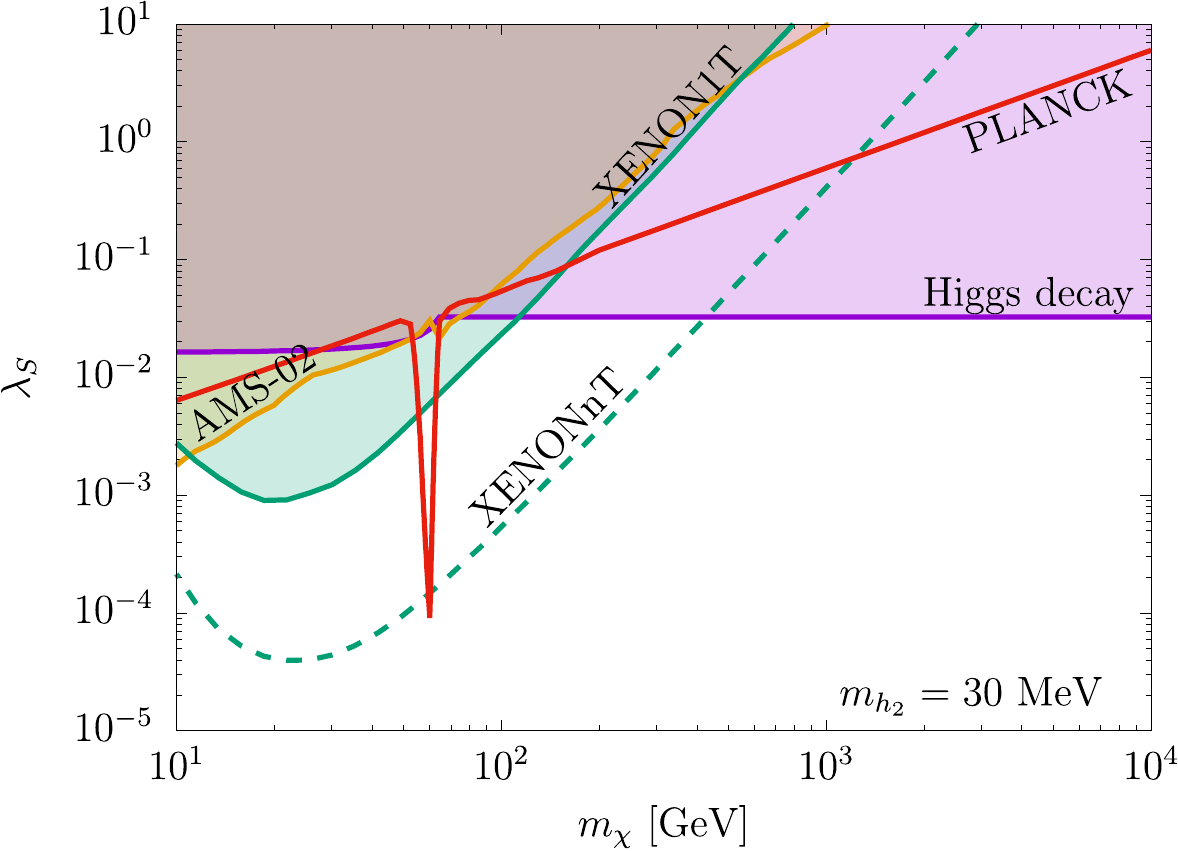}
  \includegraphics[scale=0.65]{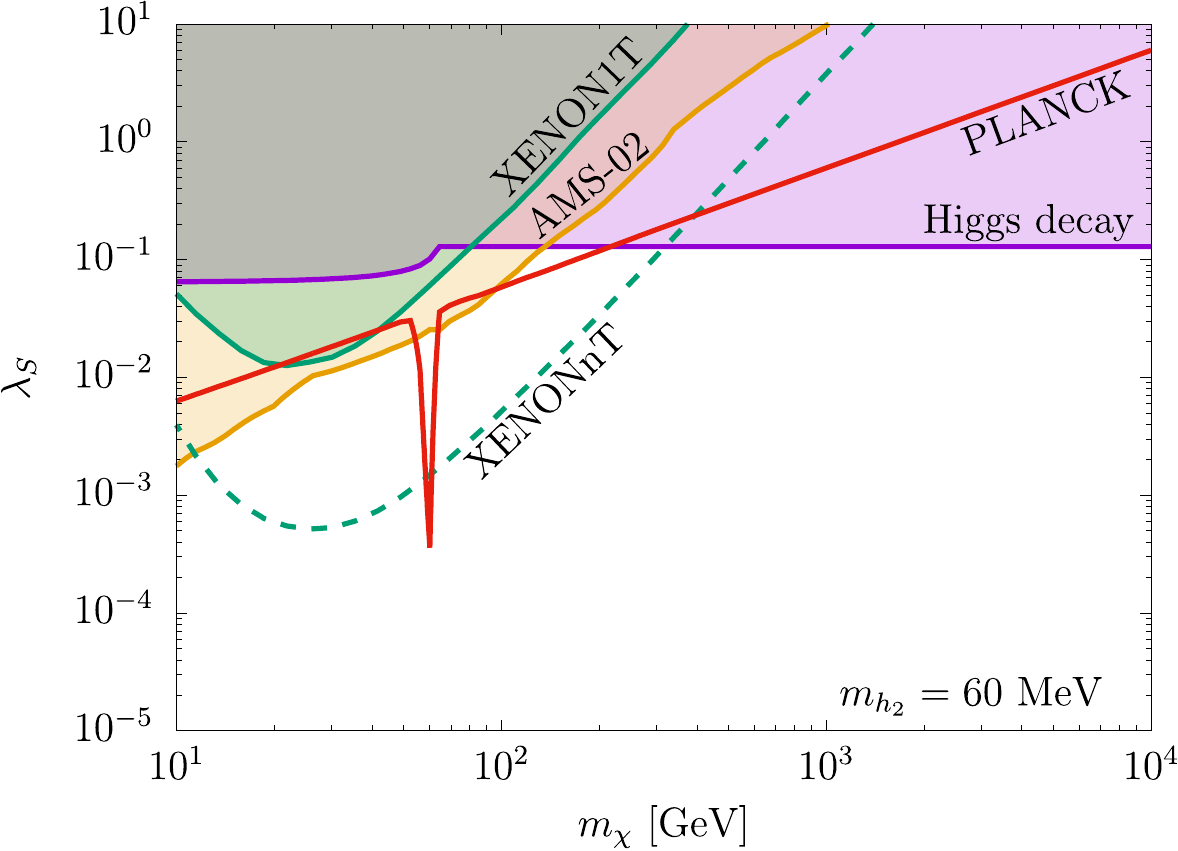}\\
  \includegraphics[scale=0.65]{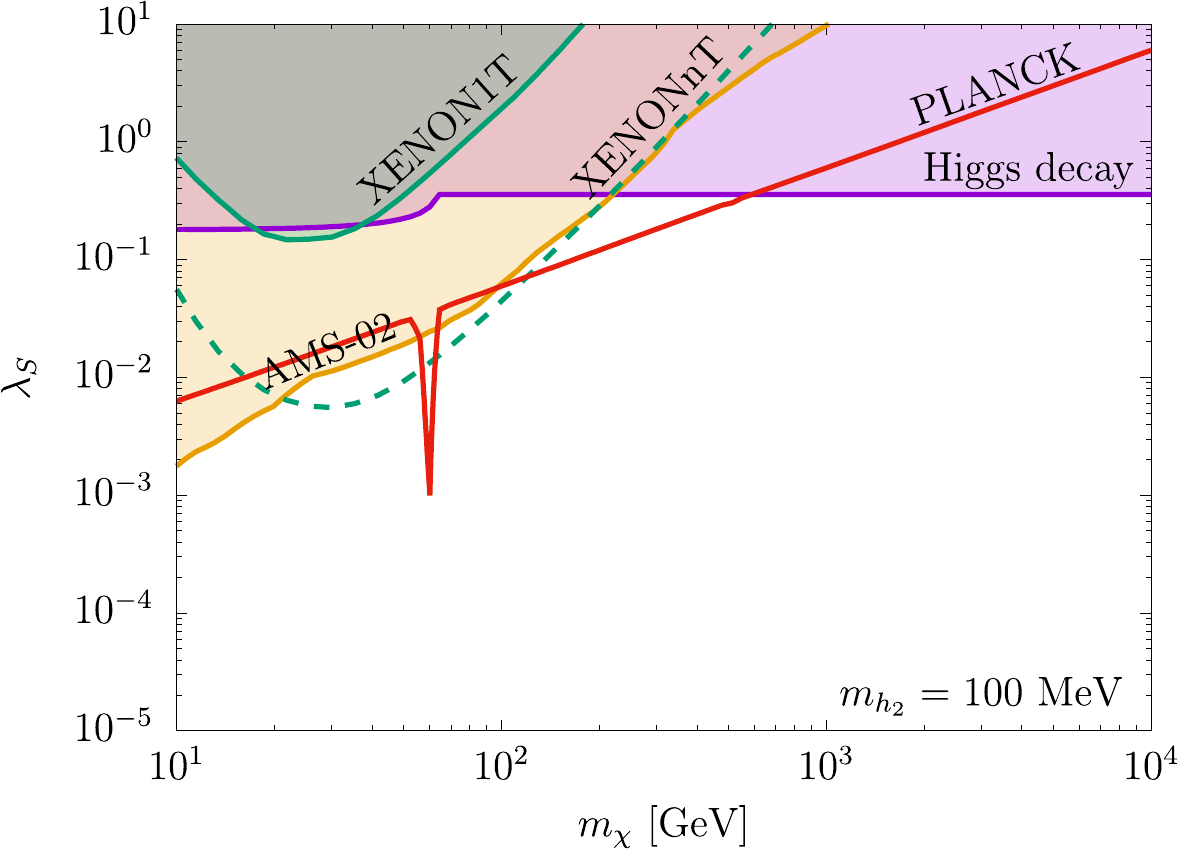}
  \includegraphics[scale=0.65]{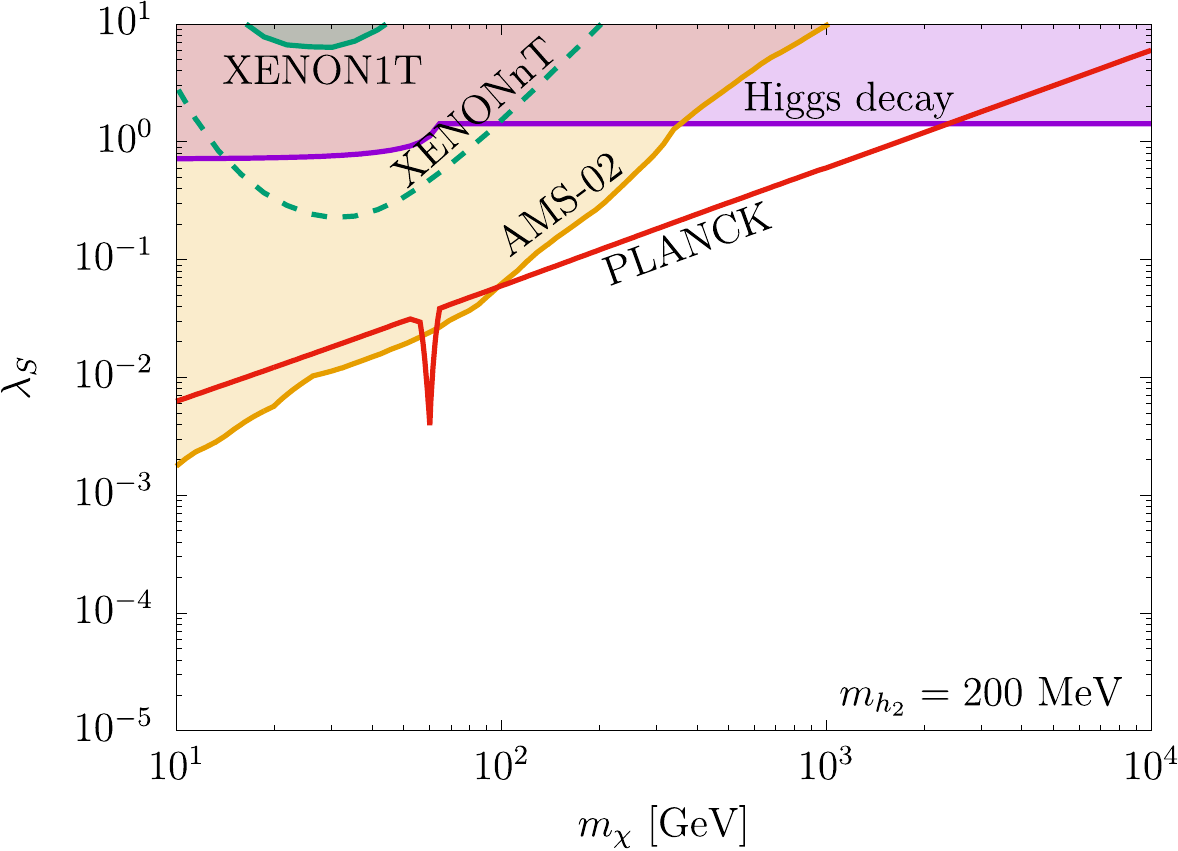}
\caption{Same plots with Fig.~\ref{fig:dd} for $\sin\theta=3\times10^{-5}$.}
\label{fig:dd2}
\end{figure}

The recoil energy spectrum for the pNGB dark matter could be discriminative from the other dark matter candidates thanks to its derivative couplings. 
The typical energy spectrum for the pNGB dark matter is shown as violet lines in the right panel of Fig.~\ref{fig:dd0}. 
The spectrum for the singlet scalar dark matter (an example of usual WIMPs) is also shown as green lines for comparison~\cite{GAMBIT:2017gge}. 
The event rate for the WIMP is enhanced at low energy while that for the pNGB dark matter is suppressed as in the figure. 
Note that WIMPs may also be able to induce energy spectra similar to the pNGB dark matter if the WIMP mass is heavier and detector efficiency is taken into account 
because the spectra can have a longer tail at higher energy for heavier WIMPs.\footnote{Similar spectra may also be induced from the other dark matter scenarios such as WIMPs with momentum dependent cross sections and inelastic scattering processes~\cite{McCabe:2010zh}.}
However, the shape of the recoil energy spectrum can be complementarily utilized to discriminate the pNGB dark matter and the other candidates combining with the other experiments and observations.
Namely, for instance if the dark matter mass is inferred in narrow range by the other experiments or observations, 
one can use the recoil energy spectrum to discriminate the pNGB dark matter from the usual WIMPs at the end.

Fig.~\ref{fig:dd} shows the parameter space in the $(m_\chi,\lambda_S)$ plane where the Higgs mixing is fixed to be $\sin\theta=10^{-4}$ and the second Higgs mass is 
$m_{h_2}=30,60,100$ and $200~\mathrm{MeV}$. 
Fig.~\ref{fig:dd2} shows the same figure with $\sin\theta=3\times10^{-5}$. 
The green, orange and violet region are excluded by the limit of XENON1T~\cite{XENON:2018voc}, AMS-02~\cite{AMS:2014bun} and Higgs decays~\cite{CMS:2018yfx, ATLAS:2019cid}, respectively. 
The green dashed line represents the future sensitivity of the XENONnT experiment~\cite{XENON:2020kmp}.
The red line can reproduce the thermal relic abundance consistent with the PLANCK observation $\Omega_\chi h^2\approx0.12$~\cite{Planck:2018vyg}.
The constraints of direct detection and Higgs decay tend to be severe for lighter $h_2$ and larger mixing $\sin\theta$ as can be seen from Eqs.~(\ref{eq:hdecay}), (\ref{eq:hdecay2}) and (\ref{eq:dsde}). 
On the other hand, the AMS-02 bound does not much depend on $m_{h_2}$ except for the region close to the Higgs resonance as obvious from Eq.~(\ref{eq:h2h2}). 
As discussed in Ref.~\cite{Chen:2014ask},
the perturbative unitarity constraint imposes the upper bound on the self quartic coupling as $\lambda_{S} < 8 \pi / 3$,
however it is invariably weaker than the constraint from Higgs decay in our scenario.
From these plots, we can see that the thermally produced pNGB dark matter with the light mediator can be consistent with all the constraints when the dark matter mass is $m_\chi\sim100~\mathrm{GeV}$ for $\sin\theta=3\times10^{-5}$ and $m_{h_2}=60~\mathrm{MeV}$, and can be tested by the future XENONnT experiment.

\section{Conclusions}
In the previous works, the pNGB dark matter has been completely insensitive to dark matter direct detection experiments because the amplitude for the elastic scattering vanishes 
in non-relativistic limit.
In this paper, we have focused on the case that the particle mediating the elastic scattering is light enough, and we have found that a non-zero contribution to the amplitude emerges. 
Together with the relevant constraints such as the thermal relic abundance, $e^+e^-$, gamma ray, CMB and Higgs decays, 
we have shown the parameter region of this scenario.
Some parameter region have already been excluded by the current XENON1T experiment and some other region can be tested by the future XENONnT experiment. 
The pNGB dark matter can be discriminative from the other dark matter candidates due to complementary study of dark matter direct detection and cosmic ray observations.

\section*{Acknowledgements}
\noindent
This work was supported by JSPS Grant-in-Aid for Scientific
Research KAKENHI Grant Nos. JP20J11901 (Y.A.), JP20K22349 (T.T.). 
Numerical computation in this work was carried out at the Yukawa Institute Computer Facility.

\appendix
\section{Annihilation cross sections}
\label{app:cross-section}

In this Appendix, we give analytic expressions for all the annihilation channels for completeness.
The annihilation cross sections for $\chi\chi\to h_ih_j$ are given by
\begin{align}
\sigma_{h_1h_1}v_\mathrm{rel}&=
\frac{1}{4\pi s}\left(|a_{11}|^2+\frac{2}{v_s^4}\frac{\sin^4\theta m_{h_1}^8}{c_{11}^{+}c_{11}^{-}}\right)
\sqrt{1-\frac{4m_{h_1}^2}{s}}\nonumber\\
&\hspace{0.5cm}+\frac{\sin^2\theta m_{h_1}^4}{4\pi s^{3/2}v_s^2\sqrt{s-4m_\chi^2}}
\left(2\Re(a_{11})+\frac{\sin^2\theta}{v_s^2}\frac{m_{h_1}^4}{s-2m_{h_1}^2}\right)\log\left(\frac{c_{11}^{+}}{c_{11}^{-}}\right),\\
\sigma_{h_1h_2}v_\mathrm{rel}&=\frac{1}{2\pi s}
\left(|a_{12}|^2+\frac{2}{v_s^4}\frac{\sin^2\theta\cos^2\theta m_{h_1}^4m_{h_2}^4}{c_{12}^{+}c_{12}^{-}}\right)
\frac{\sqrt{\lambda(s,m_{h_1}^2,m_{h_2}^2)}}{s}\nonumber\\
&\hspace{0.5cm}+\frac{\sin\theta\cos\theta m_{h_1}^2m_{h_2}^2}{2\pi s^{3/2} v_s^2\sqrt{s-4m_\chi^2}}\left(
2\Re(a_{12})+\frac{\sin\theta\cos\theta}{v_s^2}\frac{m_{h_1}^2m_{h_2}^2}{s-m_{h_1}^2-m_{h_2}^2}
\right)\log\left(\frac{c^+_{12}}{c^-_{12}}\right),\\
\sigma_{h_2h_2}v_\mathrm{rel}&=
\frac{1}{4\pi s}\left(|a_{22}|^2+\frac{2}{v_s^4}\frac{\cos^4\theta m_{h_2}^8}{c_{22}^{+}c_{22}^{-}}\right)
\sqrt{1-\frac{4m_{h_2}^2}{s}}\nonumber\\
&\hspace{0.5cm}+\frac{\cos^2\theta m_{h_2}^4}{4\pi s^{3/2}v_s^2\sqrt{s-4m_\chi^2}}
\left(2\Re(a_{22})+\frac{\cos^2\theta}{v_s^2}\frac{m_{h_2}^4}{s-2m_{h_2}^2}\right)\log\left(\frac{c_{22}^{+}}{c_{22}^{-}}\right),
\end{align}
where $s$ is the Mandelstam variable, the coefficients $a_{ij}$ and $c_{ij}^{\pm}$ are given by
\begin{align}
a_{11}&=\frac{\sin^2\theta}{v_s^2}m_{h_1}^2+\frac{\sin\theta}{2v_s}\frac{\kappa_{111}s}{s-m_{h_1}^2+im_{h_1}\Gamma_{h_1}}
 -\frac{\cos\theta}{2v_s}\frac{\kappa_{112}s}{s-m_{h_2}^2+im_{h_2}\Gamma_{h_2}},\\
a_{12}&=\frac{\sin\theta\cos\theta}{v_s^2}\frac{m_{h_1}^2+m_{h_2}^2}{2}
-\frac{\sin\theta}{2v_s}\frac{\kappa_{112}s}{s-m_{h_1}^2+im_{h_1}\Gamma_{h_1}}
+\frac{\cos\theta}{2v_s}\frac{\kappa_{122}s}{s-m_{h_2}^2+im_{h_2}\Gamma_{h_2}},\\
a_{22}&=\frac{\cos^2\theta}{v_s^2}m_{h_2}^2+\frac{\sin\theta}{2v_s}\frac{\kappa_{122}s}{s-m_{h_1}^2+im_{h_1}\Gamma_{h_1}}
 -\frac{\cos\theta}{2v_s}\frac{\kappa_{222}s}{s-m_{h_2}^2+im_{h_2}\Gamma_{h_2}},\\
c_{ij}^{\pm}&=s-m_{i}^2-m_{j}^2\pm\frac{\sqrt{s-4m_\chi^2}\sqrt{\lambda(s,m_{h_i}^2,m_{h_j}^2)}}{\sqrt{s}},
\end{align}
with the kinematic function $\lambda(x,y,z)=x^2+y^2+z^2-2xy-2yz-2zx$ and the cubic couplings $\kappa_{ijk}~(i,j,k=1,2)$ are given in Eqs.~(\ref{eq:kappa1})--(\ref{eq:kappa4}). 
$\Gamma_{h_i}$ denotes the total decay width of $h_i$.

For the other channels, the annihilation cross sections are given by
\begin{align}
\sigma_{WW}v_\mathrm{rel}&=\frac{g_2^2m_W^2}{8\pi}\sqrt{s-4m_W^2}
\frac{s^{3/2}}{v_s^2}\cos^2\theta\sin^2\theta
\left[3-\frac{s}{m_W^2}+\frac{s^2}{4m_W^4}\right]
\left|\tilde{G}_{h_1}-\tilde{G}_{h_2}\right|^2,\\
\sigma_{ZZ}v_\mathrm{rel}&=\frac{g_2^2m_Z^2}{16\pi \cos^2\theta_W}\sqrt{s-4m_Z^2}
\frac{s^{3/2}}{v_s^2}\cos^2\theta\sin^2\theta
\left[3-\frac{s}{m_Z^2}+\frac{s^2}{4m_Z^4}\right]
\left|\tilde{G}_{h_1}-\tilde{G}_{h_2}\right|^2,\\
\sigma_{f\overline{f}}v_\mathrm{rel}&=\frac{m_f^2}{8\pi}\sqrt{s}\left(s-4m_f^2\right)^{3/2}
\frac{\cos^2\theta\sin^2\theta}{v^2v_s^2}
\left|\tilde{G}_{h_1}-\tilde{G}_{h_2}\right|^2,
\end{align}
where $\tilde{G}_{h_i}$ denotes the $h_i$ propagator and $\left|\tilde{G}_{h_1}-\tilde{G}_{h_2}\right|^2$ is given by
\begin{align}
 \left|\tilde{G}_{h_1}-\tilde{G}_{h_2}\right|^2=
\left|\frac{1}{s-m_{h_1}^2+im_{h_1}\Gamma_{h_1}}-\frac{1}{s-m_{h_2}^2+im_{h_2}\Gamma_{h_2}}\right|^2.
\end{align}
Since the propagator part can be simplified as
\begin{align}
\left|\tilde{G}_{h_1}-\tilde{G}_{h_2}\right|^2
\approx
\frac{m_{h_1}^4}{s^2}\frac{1}{(s-m_{h_1}^2)^2+m_{h_1}^2\Gamma_{h_1}^2},
\end{align}
when $\sqrt{s},m_{h_1}\gg m_{h_2},\Gamma_{h_1},\Gamma_{h_2}$, one can verify the above cross sections do not violate the unitarity at $s\to\infty$.


\newcommand{\arxivfont}{\rmfamily}
\bibliographystyle{utphys}
\bibliography{ref}

\end{document}